\documentclass[10pt,reqno]{amsart}
     \makeatletter
     \def\section{\@startsection{section}{1}%
     \z@{.7\linespacing\@plus\linespacing}{.5\linespacing}%
     {\bfseries%\normalfont\scshape
     \centering
     }}
     \def\@secnumfont{\bfseries}
     \makeatother
\newtheorem{theorem}{Theorem}[section]

\theoremstyle{definition}
\newtheorem{definition}[theorem]{Definition}

\theoremstyle{remark}

\numberwithin{equation}{section}
\usepackage{amssymb}
\usepackage{amsmath}
\usepackage{mathrsfs}
\usepackage{amsfonts}
\usepackage{dutchcal}
\usepackage{braket}
\usepackage{physics}
\usepackage{comment}
\usepackage{color}
\usepackage{graphicx}
\usepackage{tikz}
\usetikzlibrary{calc, shapes, graphs, positioning,backgrounds,arrows,automata,backgrounds,fit,decorations.pathreplacing,hobby,celtic,knots,decorations.pathreplacing,shapes.geometric,intersections
}

\usepackage[scr=boondox,  % heavily sloped
            cal=esstix]   % slightly sloped
           {mathalpha}

\def\XXint#1#2#3{{\setbox0=\hbox{$#1{#2#3}{\int}$ }
\vcenter{\hbox{$#2#3$ }}\kern-.6125\wd0}}

\let\origmaketitle\maketitle
\def\maketitle{
  \begingroup
  \def\uppercasenonmath##1{} % this disables uppercasing title
  \let\MakeUppercase\relax % this disables uppercasing authors
  \origmaketitle
  \endgroup
}

\definecolor{Hey}{rgb}{.9,.05,.4}
\definecolor{orange}{rgb}{1,.5,0}
\definecolor{plum}{rgb}{.4,0,.6}
\definecolor{R}{rgb}{1,0,0}
\definecolor{G}{rgb}{0,1,0}
\definecolor{B}{rgb}{0,0,1}

\begin {document}
\title{Quantum Information Approach to Bosonization of \newline Supersymmetric Yang-Mills Fields }

\author {Radhakrishnan Balu$^\dag$}
\thanks{$^\dag$Corresponding Author.}
\address{Department of Physics, 
	University of Maryland,
	College Park, MD 20742.}
\email{rbalu@umd.edu}
\address{Army Research Office Adelphi, MD, 21005-5069, USA}
\urladdr{http://sites.google.com/view/radbalu}
\vskip0.2in
\author {S.\ James Gates, Jr.}
\address{Department of Physics \& School of Public Policy, 
	University of Maryland,
	College Park, MD 20742-4111}
\email{rbalu@umd.edu}
\address{University of Maryland, College Park, MD 20742-4111, USA}
\urladdr{http://sjgatesjr@umd.edu}

\subjclass[2010] {Primary supersymmetry; Secondary Yang-Mills field theory}

\keywords{supersymmetry; Yang-Mills Fields; deformed Heisenberg algebras; bosonization.}

\begin{abstract} We consider bosonization of supersymmetry in the context of Wess-Zumino quantum mechanics. Our motivation for this investigation is the flexibility the bosonic fock space affords as any classical probability distribution can be realized on it making it a versatile framework to work with for quantum processes. We proceed by constructing a minimal bosonization of a system with one bosonic and two fermionic degrees of freedom. We iterate this process to construct a tower of SUSY systems that is akin to unfolded Adinkras. We then identify an $osp(2|2)$ symmetry of the system constructed. To build an irreducible representation of the system we induce representations across the sectors, a first to our knowledge, as the previous work have focused on induction only within the bosonic sector. First, we start with a fermionic representation using Clifford algebras and then induce a representation to $gl(2|2)$ and restrict it to $osp(2|2)$. In the second method, we induce a representation from that of the bosonic sector. In both cases, our representations are in terms of qubit operators that provide a way to solve SUSY problems using quantum information based approaches. Depending upon the direction of induction the representations are suitable for implementation on a hybrid qubit and fermionic or bosonic quantum computers.
 
\end{abstract}

\maketitle

%%\noindent $\clubsuit$ Note to author: Use 2010 Mathematics Subject Classification.
\section{Introduction}
\label{intro}
Among the most interesting developments regarding the classes of bosonic versus fermionic field operators are the existences of bosonization or fermionization transformations between the two classes.  A number of historical precedents led to this legacy.  These include works of
Jordan-Wigner \cite{X1}, Tomonaga \cite{X2}, Luther-Peschel \cite{X3},
and Mattis-Lieb \cite{X4}.  While these
transformations are often referred to as `Klein Transformations,' it seems this name was appended fairly late in the
history of these developments in quantum field theory \cite{X5}.  In this lineage
of previous works, one of the authors (SJG), considered the Klein Transformation in the context of low dimensional supersymmetric
field theories in the papers \cite{X6,X7}. \newline

Our contribution in this work has three parts. We constructed an infinite tower of SUSY systems and the q-deformed version describes SUSY breaking. We then built IRRs for the super Lie group $ops(2|2)$ using the tools of induced representations. We carried out the constructions using Mackey machinery in two directions of bosonic-to-fermionic and vice versa as the previous work, in the SUSY context, induced representations only within the bosonic sector \cite {Radbalu2024, Varadarajan2011}. These earlier inductions involved super Lie subgroups to super Lie groups with the same super Lie algebras and so only within the bosonic sector one could achieve generalization of symmetry. We related the constructions to highest weight IRRs that are ubiquitous \cite {Tol1986} in reduced Lie algebra techniques. Finally, we cast the constructions in the language of quantum information processing and indicated possible quantum computing formalisms for realizing them.

\section{Preliminaries}
We can $Z_2$ grade a bosonic fock space to describe a supersymmetrical fock space with fermionic and bosonic degrees of freedom. A Klein operator, $K$, that satisfies $$K^2 = 1; \{K, a\} = 0 = \{K, a^\dag\}$$
enables the realization of a SUSY system as it splits the fock space with the orthonormal states $\ket{n} = (n!)^{-1/2}(a^\dag)^n \ket{0}, n = 0, 1, 2, \dots ; a\ket{0} = 0 $ into odd and even spaces $K\ket{n} = \kappa (-1)^n \ket {n}$. Without loss of generality, we can assume $\kappa = +1$ and the operator can be realized using the Euler{'} formula as $K = e^{i\pi N} = cos(\pi N) + i sin(\pi N) = cos(\pi N) = (-1)^N, N = a^\dag a, N\ket{n} = n\ket{n}$. In the coordinate representation we can express the creation and annihilation operators as 
\begin {align}
a^\dag &= \frac {1}{\sqrt{2}}(x - ip). \\
a &= \frac {1}{\sqrt{2}}(x + ip). \\
p &= id/dx.
\end {align}
In this representation $K$ can be thought of as a parity operator that acts on the wavefunction as 
\begin {equation}
K\psi_\pm = \pm \psi_\pm, \psi_\pm = \frac {1}{2} (\psi(x) \pm \psi(-x)).
\end {equation}

The above bosonic oscillator can be deformed by a parameter $\nu$ that can be described the Heisenberg algebra satisfying the relation 
\begin {align}
a^\dag &= \frac {1}{\sqrt{2}}(x - ip). \\
a &= \frac {1}{\sqrt{2}}(x + ip). \\
[a, a^\dag ] &= 1 + \nu K. \\
p &= -i(\frac {d}{dx} - \frac {\nu} {2x}K).
\end {align}
With the value of $\kappa$ as above, we get this expression for the number operator operating in the states $\ket{n} = C_n (a^\dag)^n \ket{0}$ as 
$$N \ket{n} = [n]_\nu \ket{n}, [n]_\nu = n + \frac {\nu}{2}(1 + (-1)^{n + 1})$$. It follows that when $\nu > -1$ the deformed algebra has the basis $\ket{n}$ with the normalization coefficient 
\begin {equation}
C_n = ([n]_\nu !)^{-\frac{1}{2}}, [n]_\nu ! = \prod_{k = 1}^n [k]_\nu
~~~.
\end {equation}
and the number operators are expressed as 
\begin {equation} \label {eq: defNumOp}
N = \frac {1}{2} \{ a, a^\dag \} -\frac {1}{2} (\nu + 1)
~~~.
\end {equation}

The deformed annihilation and creation operators can be expressed as
\begin {equation} \label {eq: deformed ops}
a^{\pm} = \frac {1}{\sqrt{2}} (x \mp ip) \text {   }
{~~~}, {~~~~~} ~~p = i(\frac {d}{dx} - \frac {\nu} {2x}K) ~~~.
\end {equation}

Now let us consider Wess-Zumino quantum mechanics (WZQM) which has one complex bosonic variable $\phi(t) = x(t) + i y(t)$ and one complex fermionic degrees of freedom $\psi(t)_\alpha, \alpha = 1,2$. With this set up the Hamiltonian reads as 
\begin {equation} \label {eq: hamiltonian}
H = \frac {1}{2} (p_x^2 + p_y^2) + |\phi + g_0\phi^2 |^2 + ((1 + 2h_0\phi)\psi \psi^\dag + h.c.).
\end {equation}

The corresponding supercharges are:
\begin {align}
Q &= \frac {1}{\sqrt{2}} i(p + (\phi + g_0\phi^2))\psi + (\phi + g_0\phi^2))\psi^\dag .\\
Q^\dag &= \frac {1}{\sqrt{2}} i(p - (\phi + g_0\phi^2))\psi^\dag + (\phi + g_0\phi^2))\psi .\\
W(x) &= (\phi + g_0\phi^2). \\
\{\psi, \psi^\dag\} &= 1. \\
\psi^2 &= 0. \\
(\psi^\dag)^2 &= 0. \\
QQ^\dag &= \frac {1}{2}\psi \psi^\dag (p^2 + i[p, W'(x)] + (W'(x))^2). \\
Q^\dag Q &= \frac {1}{2}\psi^\dag \psi(p^2 -i[p, W'(x)] + (W'(x))^2) \\
& \text{   as  [p, x] = -i and [p, f(x)] = -idf(x)/dx.} \\
\psi &= a^- K. \\
\psi^\dag &= a^+ K.
\end {align}

One way to construct a spontaneous SUSY breaking system is to compose an odd operator and projectors $\Pi_\pm = \frac {1}{2} (1 \pm K)$ \cite {Mikhail1996}. This follows from the fact that the operator $1 + K$ annihilates odd states, and when composed with an odd operator, we get a nilpotent supercharge. More formally, 

\begin {align*}
(Q\Pi_\pm)^2 &= Q\Pi_\pm Q\Pi_\pm. \\
&= -Q^2 \Pi_\pm. \\
&= 0.
\end {align*}

As the above $Q$ operators are odd, we can set up new supercharges as $\hat{Q} = Q \Pi_\pm$ for an $N = 2$ SUSY system. The new Hamiltonian will have additional terms as;
\begin {align*}
2\hat{H} &= \{ \hat{Q}^*, \hat{Q} \}. \\
2\hat{H} &= Q(K + 1)Q^\dag (1 - K) + Q^\dag (1 - K)Q(1 + K), \\
&= -QKQ^\dag K  +  QKQ^\dag - QQ^\dag K + QQ^\dag - Q^\dag KQK - Q^\dag KQ + Q^\dag QK + Q^\dag Q. \\
&= QQ^\dag + Q^\dag Q + (-QQ^\dag + Q^\dag Q)K  +  (-QKQ^\dag - Q^\dag KQ)K + QKQ^\dag - Q^\dag KQ.\\
&= QQ^\dag + Q^\dag Q + (-QQ^\dag + Q^\dag Q)K  +  (QQ^\dag + Q^\dag Q) + (-QQ^\dag + Q^\dag Q)K. \\
&= H + [Q^\dag , Q]K.
\end {align*}

Replacing $K$ with $K + 1$ and $K - 1$ in the Hamiltonian (equation \eqref {eq: hamiltonian}) we get two SUSY systems for WZQM. Now, our fermionic variables are $a^\pm (K \pm \mathbb{I})$ and it is easy to verify that they satisfy the fermionic relations. Let us look at their spectrum and determine whether SUSY is spontaneously broken.
The original WZQM Hamiltonian has a unique zero-energy ground state, that is, it is an exact SUSY system. So, $a^- Ka^+ K$ has to be zero for the $n = 0$ state only. In the new Hamiltonian, there are three additional terms for each of the two SUSY systems. Let us first consider the case of $K + 1$ and the last term involving the fermionic variables at the Hamiltonian has;
$a^- Ka^+ K + 1 + a^- K + a^+ K$. This has a value of $1$ at $n = 0$ and again $1$ at $n = 1$ providing degeneracy and thus spontaneous SUSY breaking. 

We can reason similarly for the $K - 1$ case and find that SUSY is again spontaneously broken. Using the deformed algebra, we can change the scale of SUSY breaking.

The $\nu$-deformed version of the Hamiltonian will have the $p$ term suitably modified by the $\nu$ parameter (equation \eqref{eq: deformed ops}. Fermionic operators are derived from the Jordan-Wigner transformation and deformation applied to the $K$ term as $1 \pm K$ and using \eqref{eq: defNumOp}. Now, we have two SUSY systems, and the one corresponding to $1 - K$ will lead to spontaneous symmetry breaking with different degrees.

This construction generalizes easily to Hamiltonians with multiple bosonic or fermionic variables as we can take the corresponding supercharges and create new ones using the Klein projectors. In fact, this construction can be repeated infinite times and each time we get an additional commutator term in the Hamiltonian. With the aid of a deformed harmonic oscillator we can repeat the construction increasing the degree of SUSY breaking with each iteration.

\section {$OSp(2|2)$ supersymmetric YM systems}
Again, following the constructions of \cite {Mikhail1996} we set up the even and odd generators of the supergroup $OSp(2|2)$ as:
\begin {equation} \label {eqref: Osp(2|2) YM}
\begin {aligned}
T_3 &= \frac{1}{2} (a^+ a^- + a^- a^+). \\
T_{\pm} &= (a^{\pm})^2. \\
J &= -\frac{1}{2} \epsilon K [a^-, a^+]. 
\end {aligned}
\begin {aligned}
Q^{\pm} &= Q^{\mp}_\epsilon. \\
S^{\pm} &= Q^{\mp}_{-\epsilon}. 
\end {aligned}
\end {equation}
It is easy to verify that these operators satisfy the commutator relationship of the $OSp(2|2)$ superalgebra. This description of the super YM system covers both exact and broken SUSY depeding upon whether the parameter $\epsilon$ is $\pm$. However, to make this useful, that is to build the larger symmetry (exact SUSY) from the smaller one (broken SUSY) we need the representation theory of Lie super algebras. We would like to do this because we can construct the spectrum of the larger system from that of the smaller one using branching rules or intertwiners (generalized Clebsch-Gordon coefficients). In other words, from the spectrum of the fermionic (or bosonic) sector of the system we can compute all the energy levels of the SUSY system.

 Our goal here is not to describe the formation of supersymmetry, rather the computation of SUSY representations from that of the simple symmetric representations. 
 %We compute the reduction algebra and the extremal projector as outlined above and using the highest weight representation of the Clifford algebra, which is the spinor representation with action of Pauli matrices on $\mathbb{C}^2$, and determine the cyclic vector of a representation of the bigger $gl(1|2)$ algebra. Let $h$ be the cyclic vector of $V$ whc is is the representation space for $gl(1|2)$ and $V_+$ is the spinor space of the Cifford representation, then we have this relation $$V_+ = Z(gl(1|2), osp()|2) h.$$ As the extremal projector is expressed in terms of $gl(1|2)$ we can compute $h$ and thus the space $V$.

\section {Systems of Imprimitivity}
Mackey machinery to construct systems of imprimitivity (SI) constitute a comprehensive set of tools to characterize the unitary representations of Lie groups.  SI for a locally compact group (that includes compact groups) such as Poincar\'e is a composite object $(G, \Omega)$ of a representation of a group $G$ and its action on a $G$-space $\Omega$ which is usually the configuration space of the system under consideration. We denote the system of imprimitivity $(G, \Omega)$ lives on the $G-$orbit $\Omega$. Mackey machinery enables induction of irreducible representations of a group $G$, from that of a subgroup $H$, that are systems of imprimitivity. The configuration spaces of interest to us in this work are orbits of Pauli and Clifford groups and the homogeneous space $G/H$, where $H$ is a closed subgroup of $G$ that consists of left cosets $gH, g \in G$. From SI characterizations, we can derive the canonical commutation relations and infinitesimal forms in terms of equations (Schr$\ddot{o}$dinger, Heisenberg, and Dirac etc) \cite {Wigner1949} \cite {Rad2019} and \cite {Rad2024}. In this work, we will apply the SI techniques to build Heisenberg-Weyl operators on super Hilbert spaces and then construct a representation for the super Lie group $osp(2|2)$. In our earlier work, we have constructed covariant Quantum Fields via Lorentz Group Representation of Weyl Operators \cite {RadBalu2019, RadBalu2020}. Here, we specialize the techniques for compact Lie groups and in particular to $osp(2|2)$.

\section {Stabilizer subgroups (Little groups)}
In this section we describe some examples of systems of imprimitivity that live on the orbits of the stabilizer subgroups of Poincar\'e. It is good to keep in mind the picture that SI is an irreducible unitary representation of Poincar$\grave{e}$ group $\mathscr{P}^+$ induced from the representation of a subgroup such as $SO_3$. This is a subgroup of homogeneous Lorentz, as $(U_m(g)\psi)(k) = e^{i\{k,g\}}\psi(R^{-1}_mk)$ where g belongs to the $\mathscr{R}^4$ portion of the Poincar$\grave{e}$ group, m is a member of the rotation group. The duality between the configuration space $\mathbb{R}^4$ and the momentum space $\mathbb{P}^4$ is expressed using the character the irreducible representation of the group $\mathbb{R}^4$ as:
\begin {align} \label {eq: poissonBr}
\{k,g\} &= k_0 g_0 - k_1 g_1 - k_2 g_2 - k_3 g_3, p \in \mathbb{P}^4. \\
\hat{p}:x &\rightarrow e^{i\{k,g\}}. \\
\{Lx, Lp \} &= \{ x, p \}. \\\
\hat{p}(L^{-1}x) &= \hat{Lp}(x).
\end {align}
In the above, $L$ is a matrix representation of homogeneous Lorentz group acting on $\mathbb{R}^4$ as well as $\mathbb{P}^4$ and it is easy to see that $p \rightarrow Lp$ is the adjoint of L action on $\mathbb{P}^4$. The $\mathbb{R}^4$ space is called the configuration space and the dual $\mathbb{P}^4$ is the momentum space of a relativistic quantum particle.
\\

The stabilizer subgroup of the Poincar$\grave{e}$ group $\mathscr{P}^+$ (Space-like particle) can be described as follows: \cite {Kim1991}:
The Lorentz frame in which the particle is at rest has momentum proportional to (0,1,0,0) and the little group is
again SO(3, 1) and this time the rotations will change the helicity. In a similar fashion, in this work we are interested in subgroups of Pauli group that have stability points (states). In the Lie algebraic settings they play the role of Cartan subalgebras or maximal tori.
\
\begin {definition} \label {GradedHS} A super Hilbert space is a $Z_2$-graded super vector space $\mathscr{H} = \mathscr{H}_1 \oplus \mathscr{H}_2$ over $\mathbb{C}$ with a scalar product $(. , .)$, where the $\mathscr{H}_i (i = 0, 1)$, referred as even and odd, are closed mutually orthogonal subspaces. We set up the parity operator as $$ p(x) = \begin{cases} 0, & \text{if x }\in \mathscr{H}_0, \\ 1, & \text{if x }\in \mathscr{H}_1. \end {cases}$$ We define an even super Hilbert form
$$\langle x,y\rangle = \begin{cases} 0, & \text{if x and y are of opposite parity} \\ (x,y), & \text{if x and y are even} \\ i(x,y), & \text{if x and y are odd} \end {cases}.$$
We have $$\langle y,x \rangle = (-1)^{p(x)p(y)}\overline{\langle x,y \rangle}.$$
\end {definition}
If $T(\mathscr{H} \rightarrow \mathscr{H})$ is a bounded linear operator, we denote by $T^*$ its Hilbert space adjoint and by $T^{\dag}$ its super adjoint given by $\langle Tx, y \rangle = (-1)^{p(T)p(x)}\langle x, T^{\dag} y \rangle$. Here, $p(T^\dag = p(T)$ and the parity of the operator as $T$ is even or odd.

\begin {definition} A super Lie group is $(G_0, \mathcal{g})$ is a super Harish-Chandra pair if $G_0$ is a classical Lie group and $\mathcal{g}$ is a super Lie algebra with an action of $G_0$ on it such that
(i) Lie($G_0)$) = $\mathcal{g}_0$ = the even part of $\mathcal{g}$.
(ii) The action of $G_0$ on $\mathcal{g}$ is the adjoint action of $G_0$; more precisely, the adjoint action of $G_0$ on $\mathcal{g}$ is the differential of the action of $G_0$ on $\mathcal{g}$.
A representation of a super Lie group is a triplet ($\pi, \gamma, \mathscr{H}$), where $\pi$ is an even representation of $G_0$ in a super Hilbert space $\mathscr{H}$ and $\gamma$ is a super representation of $\mathcal{g}$ in $\mathscr{H}.$
\end {definition}

\begin {definition} A super Lie algebra is a super vector space $\mathcal{g}$ with a bilinear bracket $[ , ]$ such that $\mathcal{g}_0$ is an ordinary Lie algebra with $[.,.]$ and $\mathcal{g}_1$ is a $\mathcal{g}_0$-module for the action $a \rightarrow ad(a): b \rightarrow [a, b], (b \in \mathcal{g}_1$). Further, $a \otimes b \rightarrow [a, b]$ is a symmetric $\mathcal{g}_0$-module map from $\mathcal{g}_1 \otimes \mathcal{g}_1$ to $\mathcal{g}_0$. It also satisfies the nonlinear condition $$[a, [a, a]] = 0, \forall g \in \mathcal{g}_1.$$ One way, to ensure this last condition is met, is to require that the range of the odd bracket $\mathcal{g}_2$ is a subset of $\mathcal{g}_0$ which acts on $\mathcal{g}_1$ trivially as $$\mathcal{g}_2 \subset \mathcal{g}_0 \Rightarrow [\mathcal{g}_1, \mathcal{g}_0] = 0.$$ A super algebra $A$ is an algebra of endomorphisms of linear maps on a super vector space $V$. The maps that preserve he grading of $V$ are designated as even and those reverse it are called odd. To get a super Lie algebra from $A$ we ca use the bracket $$[a, b] = ab - (-1)^{p(a)p(b)}ba.$$

%which is a morphism  $\mathcal{g} \otimes \mathcal{g} \rightarrow \mathcal{g}$ with the following properties:
%\begin {align*}
%[a, b] = -(-1)^{p(a)p(b)}&[b, a]. \\
%[a, [b, c]] + (-1)^{p(a)[p(b)+p(c)]}[b, [c, a]] + %(-1)^{p(c)[p(a)+p(b)]}&[c, [a, b]] = 0.
%\end {align*}
% The (super) Jacobi identity (unscrambling the earlier version in a tensor category)
\end {definition}
Let us recollect the notions of systems of imprimitivity (SI) and an important result by Mackey that characterizes such systems in terms of induced representations, key notions in Clifford algebras, spinor fields, and Schwartz spaces \cite {Varadarajan1985}, before discussing our main result in the super context. We provide the SUSY generalizations along with their classical counterparts using the notations and notions from the works of Varadarajan \cite {Varadarajan1985, Varadarajan2006, Varadarajan2011}.

\begin {definition} \cite {Varadarajan1985} A G-space of a Borel group G is a Borel space X with a Borel automorphism $\forall{g\in{G}},t_g:x\rightarrow{g\cdot x},x\in{X}$ such that
\begin {align}
&t_e \text{ is an identity} \\
&t_{g_1,g_2} =t_{g_1}t_{g_2}.
\end {align}
The group G acts on X transitively if $\forall{x,y\in{X}},\exists{g}\in{G} \text { such that } {x=g\cdot y}.$
\end {definition}
\begin {definition} \cite {Varadarajan1985} A system of imprimitivity, for a group $G$ acting on a Hilbert space $\mathscr{H}$, is a pair (U, P), where $P: E\rightarrow{P_E}$ is a projection valued measure defined on the Borel space X with projections defined on the Hilbert space and U is a representation of G satisfying
\begin {equation}
U_gP_EU^{-1}_g = P_{g\cdot E}
\end {equation}  
\end {definition} 
Systems may be decomposed into SI $(G_0, \Omega = G_0/H_0)$, where $H_0$ is a closed subgroup (for example a stabilizer subgroup) of $G_0$, and a stabilizer at $\omega_0 \in \Omega$ on orbits by the transitive actions of the group and there exists a functor between the category of unitary representations of $H_0$ and the category of SI $(G_0, \Omega)$. In the case of Poincar\'e group, with stabilizer subgroups as the three little groups with constant momentum in a reference frame, a transitive SI is of interest to us and so we use the specialized version of the Mackey machinery. If $\sigma$ be a representation of $H_0$ on a Hilbert space $\mathcal{K}^\sigma$, then there is a canonical SI $(\pi^\sigma, P^\sigma)$ for $G_0$ based on $\Omega$ with the representation induced by that of $H_0$ and the natural projection valued measure on $\mathcal{K}^\sigma$. The Hilbert space is the set of equivalence classes of measurable functions $f: G_0 \rightarrow \mathcal{K}^\sigma$ satisfying:
\begin {align}
f(x\eta) &= \sigma(\eta)^{-1}f(x), \text{ for almost all }\eta \in H_0. \\
\int \abs{f(x)}^2_{\mathcal{K}^\sigma}dx & < \infty.
\end {align}
The representation $\pi^\sigma$ acts by left translations and the SI relation $\sigma \rightarrow (\pi^\sigma, \mathcal{K}^\sigma)$ states that there is a functor that exists between the category of unitary representations of $H_0$ and the category of SI based on $\Omega$. One can develop an intuition \cite {Varadarajan2011} for this construction as the Hilbert space $\mathcal{K}^\sigma$ is attached to the fixed point $\omega_0$. For all the non-fixed points $\omega = g[\omega_0]$ a Hilbert space $\mathcal{K}^\sigma_\omega$ is associated via an unitary isomorphism. This results in a fiber bundle $\mathcal{V}^\sigma = \mathcal{K}^\sigma \times G_0/ \sim$, where the equivalence relation is defined by $(g, \psi) \sim (g\eta, \sigma(\eta)^{-1}\psi)$. The group $G_0$ has a a natural right action on the bundle.
//

For constructing the irreducible representations of $osp(2|2)$ we proceed by invoking the standard version of systems of imprimitivity techniques that use representations of closed subgroups to induce representations of a larger group. This means, we induce a representation of \cite{Radbalu2024} $$Ind_{osp(0|2)}^{osp(1|2)}:osp(0|2) \rightarrow osp(1|2).$$
%by using Morita equivalence maps .
We can repeat the process to get the representation for $osp(2|2)$, though we will not detail the steps here. More precisely, we induce a representation of $$Ind_{osp(0|2)}^{gl(1|2)}:osp(0|2) \rightarrow gl(1|2)$$ and then use the fact that the group $osp(n|m)$ embeds in the super Lie group $gl(n|m)$. When an IRR is restricted to a subgroup the resulting representation may not be irreducible. Here, the representations (fundamental) are on the same base Hilbert space (carrier space of the representation) of $\mathbb{C}^2 \otimes_s \mathbb{C}^2 \oplus \mathbb{C}^2 \otimes_a \mathbb{C}^2$ and so an IRR of $gl(2|2)$ when restricted to $osp(2|2)$ will still be an IRR. We accomplish this generalization of symmetry by reversing the process described above. That is, we start with a representation of the smaller algebra $osp(0|2)$ and increase the dimension by $1$ using the matrix unit $e_ii$ and construct the larger algebras as previously. 

Let us now state and discuss the main result for the case of $osp(2|2)$ with Clifford group by constructing a strict cocycle from the representation of a subgroup following the prescription (lemma 5.24) in Varadarajan{'}s text. The SI is a consequence of strict cocycle property and  the construction is not canonical. This construction enables the application of quantum information processing techniques to the supersymmetric context. That is, starting from qubit system we can generate a representation of the group $osp(2|2)$. Alternately, we can start from the bosonic qubits with Pauli group as the symmetry and construct a representation for the orthosymplectic group.

Although, we chose to use two-qubit representations to make the actions on the bosonic and fermionic sectors explicit. This means, instead of defining the action for $g \in G$ we define the action $ (g, g) \in G \times G$ as the constructions easily generalizes to n-qubit representations on the b\'{e}b\'{e} (finite dimensional) fock space.

\begin {definition}
The Pauli group $\mathscr{P}_1 = gl(2|0)$ has the generators $\mathbb{I}_2, \pm i, \pm 1$ and 
$$
X = \begin{pmatrix} 0 & 1 \\ 1 & 0 \end{pmatrix}, Z = \begin{pmatrix} 1 & 0 \\ 0 & -1 \end{pmatrix}, Y = \begin{pmatrix} 0 & -i \\ i & 0 \end{pmatrix}.
$$
The generalized version is $\mathscr{P}_n = \langle i\mathbb{I}, X, Y, Z \rangle ^{\otimes n}.$
The Clifford group corresponding to the namesake algebra $\mathscr{C}_1 = gl(0|2)$ with the generators 
$$
H = \frac{1}{\sqrt{2}}\begin{pmatrix} 1 & 1 \\ 1 & -1 \end{pmatrix}, S = \begin{pmatrix} 1 & 0 \\ 0 & i \end{pmatrix} .
$$
The generalized version is $\mathscr{C}_n = \{ U | U g U^\dag \in \mathscr{P}_n, \forall g \in \mathscr{P}_n \}.$ \\

Matrix units: 
\begin {align*} \label {eq: MatrixUnits}
\mathscr{M}_{ij} &= \ket{e_i}\bra{e_j}; \text { Matrix with a 1 at the ij position}. \\
A_n &= \mathbb{I}_{n - 1]} \otimes \ket{e_0}\bra{e_1} \otimes \mathbb{I}_{[n + 1}. \\
A_n^\dag &= \mathbb{I}_{n - 1]} \otimes \ket{e_1}\bra{e_0} \otimes \mathbb{I}_{[n + 1}. \\
A_n A_n^\dag &= \mathbb{I}_{n - 1]} \otimes \ket{e_0}\bra{e_0} \otimes \mathbb{I}_{[n + 1}. \\
\Lambda_n &= A_n^\dag A_n = \mathbb{I}_{n - 1]} \otimes \ket{e_1}\bra{e_1} \otimes \mathbb{I}_{[n + 1} \text {  Number operator}. 
\end {align*}
The matrix units form the canonical basis the Pauli operators can be defined in terms of them and the stability states remain the same.

Each of the four Bell states are stabilizer states of the Pauli subgroup $\{Z_1 Z_2, X_1 X_2\}$ that satisfies the relation $$[Z_1 Z_2, X_1 X_2] =0$$
making it a commutative one. This is a possible example of a subgroup to induce representations as detailed below.
\end {definition}

\begin {theorem} Representation of the super group $gl(2|2)$ on the Hilbert space space $\mathbb{C}^2 \otimes_s \mathbb{C}^2 \oplus \mathbb{C}^2 \otimes_a \mathbb{C}^2$ is a transitive system of imprimitivity living on the orbit $\Omega_2 \subset \mathbb{C}^2 \otimes \mathbb{C}^2 \oplus \mathbb{C}^2 \otimes_a \mathbb{C}^2$ (antisymmetric tensor) induced from a representation of the Clifford subgroup $\mathscr{C}_2$. %Two bosonic variables are represented on $\mathbb{C}^2$ and one fermionic variable is represented on $\mathbb{C}^2 \otimes \mathbb{C}^2$.
\end {theorem}

\begin {proof}
We start with this equivalence $\mathscr{gl}(0|2) \simeq \mathscr{sl}(2) \oplus \textit{k}\mathbb{I}, \textit{k} \in \mathbb{C}$ and consider the corresponding Clifford algebra and its spinor representation $SP$. In this representation the Pauli operators act on the complex anti-symmetric space. As a compact Lie group is completely characterized by its Lie algebra we use these two notions interchangeably. Instead of using the matrix units, that are canonical basis for the $gl$ group, for the bosonic operators we use the Pauli operators.\\

The Pauli subgroup $PA$ of $\mathscr{P}_2$ with the fermionic (to distinguish from the bosonic ones) stabilizer states as the stability points is a closed subgroup of $gl(1|2)$ with the representation $SP$. The usual SI techniques, as opposed to the super version of SI, allow us to set up the fiber bundle across the fermionic-bosonic sectors as:
\begin {align}
FB^1 &= \{(v, p) \in \mathbb{C}^2 \oplus \mathbb{C}^2 \otimes_a \mathbb{C}^2, v \in \mathbb{C}^2, p \in \Omega_1 \otimes \mathbb{C}^2 \subset \mathbb{C}^2 \otimes_a \mathbb{C}^2, \\
& H(v, p) = 0, \text { H is a SUSY Hamiltonian.} \\
&\Omega_1 = S \times S-\text { orbits of fermionic stabilizer states}. \\
\pi: &(v, p) \rightarrow p.
\end {align}
and we can define the fibers as $\mathbb{C}^2$ at a stability point $x$ and isomorphic fibers at the x-orbit of $PA_2$. The $Z_2-$graded fiber bundle has the polarity and inner product defined according to Definition \ref {GradedHS}. 
 We take the sections $\phi: p \rightarrow \phi(p)$ of the fiber as the Hilbert space for the representation. The inner products of the super Hilbert space $\mathbb{C}^2 \oplus \mathbb{C}^2 \otimes_a \mathbb{C}^2$ are defined on the bosonic and fermionic sectors in the usual way. For example, the norm is defined as:
 \begin {align*}
 \norm{\phi(p)}^2 &= \int_\Omega p_0^{-1} \langle \phi(p), \phi(p) \rangle d\alpha(p), \\
 &\text { where }\alpha \text{ is the Haar measure of the compact Lie group}.
 \end {align*}
\begin{comment}
In the language of the above lemma, the homomorphism $m$ is the map 
\begin {align*}
m: &P_n \rightarrow \omega_1 \otimes \mathbb{C}^2. \\ 
b(g) &= m(\phi(g) \text { m is the representation of the subgroup, that is Pauli operators acting on the complex space}. \\
F(g_1, g_2) &= b(g_1, g_2)b(g_2)^{-1} = m(g_1)m(g_2)b(g_2)^{-1} . \\
&= m(\phi(g_1) g_2)b(g_2)^{-1} \text { now, we have the stict cocycle in terms of the subgroup representation}. 
\end {align*}
\end{comment}
\\

The FB is covariant with respect to the operators $\mathscr{gl}(1|2)$ by transforming $(v, p)^{(g|h)}$ as $(\mathscr{P}_2(g)^{-1} v, SP_2 \otimes SP_2(h) p)$.
Then, the unitaries representing $\mathscr{gl}(1|2)$ are
$$U_{g, h}(p) = \phi(SP_2^{-1}(h) p)^{(g|h)}$$ and the irreducible representation is an SI living on $\Omega_1 = \text { Orbits of the stabilizer states}$. 
\\

We can repeat the step from going to one level higher by using again $\mathbb{C}^2$ as the fiber on the new bundle and taking the symmetric tensor with the fiber of the previous bundle. We now have several choices in selecting the stabilizer points of the bosonic Pauli operators to induce the representations from the subgroups generated by $\mathscr{P}_1$. Let us detail the representation induced from the subgroup generated by $X$ as:
\begin {align}
FB^2 &= \{(w, (v, p)) \in \mathbb{C}^2 \otimes_s \mathbb{C}^2 \oplus \mathbb{C}^2 \otimes_a \mathbb{C}^2, w \in \mathbb{C}^2, v \in \Omega_2 \subset \mathbb{C}^2, \\
& H(w, (v, p)) = 0, \text { H is a SUSY Hamiltonian.} \\
&\Omega_2 = X \times X-\text { orbits of the bosonic stabilizer states} \otimes \Omega_1\\
\pi: &(v, p) \rightarrow p.
\end {align}
Again, the FB is covariant with respect to the operators $\mathscr{gl}(2|2)$ by transforming $(w, (v, p))^{(f, g|h)}$ as $(\mathscr{P}_1(f)^{-1} w, (\mathscr{P}_1(g)^{-1} v, SP_2 \otimes SP_2(h) p))$.
Then, the unitaries representing $\mathscr{gl}(2|2)$ are
$U_{f, g, h}(p) = \phi(SP_2^{-1}(h) p)^{(f, g|h)}$ is an SI living on $\Omega_1 \oplus \Omega_2$. 
\\

The creation and annihilation operators of the bosonic sector can be constructed using the Pauli $X$ and $Y$ operators, that is by reversing coordinate representation discussed above.
\end {proof}

Now, we proceed in the other direction by starting from a representation of the Heisenberg-Weyl group and induce a representation. This time we use a b\'{e}b\'{e} fock space and the matrix units basis to construct the annihilation, creation, and number operators so that we can get bosonized representations of the system.

\begin {theorem} Bosonized representation of the group $gl(2|2)$ on the b\'{e}b\'{e} focke space $\bigoplus_{i = 1}^n \mathscr{H}^{2^{\otimes_s^i}}$,  the Hilbert space $\mathscr{H}$ will be defined as part of the proof, $Z_2$ graded by the Klein operator $K^i = cos 2i\pi \Lambda_i$, is a transitive system of imprimitivity on the orbit of a selected stability state and induced from a representation of the Pauli subgroup $\mathscr{P}^n$ on the bosonic sector. 
\end {theorem}

\begin {proof}

As before, let us consider the subgroup group of Pauli group $\mathscr{P}_n$ with stability points (states)  that is a closed subgroup of $gl(2|2)$. This time, we will induce the representation in one step as $gl(2|0) \rightarrow gl(2|2)$.  The SI machinery allows us to set up the base of the fiber bundle as: 
\begin {align}
FB &= \{(v, p) \in K^n (\mathbb{C}^{2^{\otimes_s^n}}), v \in \text { fermionic sector because of Klein operator} \}, \\
& p \in \Omega, \text { in the orbits of the selected stability point } \\
& H(v, p) = 0, H \text { is a SUSY Hamiltonian.} \\
\pi: &(v, p) \rightarrow p.
\end {align}
and we can define the fibers as $\mathbb{C}^{2{\otimes_s^j}}$ and compose the Klein operator with the Hamiltonian to get the fermionic and bosonic sectors. The $Z_2-$graded fiber bundle $FB^i$ has the polarity and inner product defined according to Definition \ref {GradedHS}. The $P_i$ unitaries act on the fibers of the bundle. We take the sections of the fiber bundle as the Hilbert space $\mathscr{H}^i$ for the representation with the action of entire $gl(2|2)$ defined on the b\'{e}b\'{e} fock space $\oplus_{i = 1}^n \mathscr{H}^{2^{\otimes_s^i}}$. 
The fiber bundles $FB^i$ are covariant with respect to the operators $\mathscr{gl}(2|2)$ by transforming $(v, p)^{(g|h)}$ as $(\mathscr{P}_i(g)^{-1} v, \mathscr{P}_i(h))$.
\\

Then, the unitaries representing $\mathscr{gl}(2|2)$ are
$U_{g, h}(p) = \phi(\mathscr{P}_i^{-1}(h) p)^{(g|h)}$ is an SI living on $\Omega = \text { Orbits of the stabilizer states}$.
\end {proof}
\
Several remarks are in order based on the above results:
\begin{enumerate}
    \item In the above using the subspace $\mathbb{S}^2 \subset \mathbb{C}^2$ will describe the qubit representations. The qubit normalization constraints can be incorporated while defining the bundle instead of the plain cross product.
    \item The fiber at the stability point, and thus at all the points in the orbit, is an one dimensional space because of the choice of our point. These details are important while designing quantum circuits for implementation of the bundle.
    \item For a specific SUSY with a Hamiltonian \eqref{eq: hamiltonian} this relationship can be encoded in defining the bundle.
    \item We used the antisymmetric and symmetric tensor products to describe the bosonic and fermionic qubits respectively hiding the details such as the Jordan-Wigner transformations if a fermionic quantum computer is not used.
    \item We can generate an IRR by disjoint union of of multi-particle sector as $$\bigoplus_{n \ge 0} \mathscr{P}_n \oplus \mathscr{C}_n.$$
    \item With the aid of the constructions discussed earlier and using the Klein operator $K$ we have embedded the fermionic fock space into the bosonic one.
    \item As the Clifford group is a normalizer of Pauli group (that is it embeds in it), the above theorems can be combined into single one using bosonization. As the operators of both sectors are the usual qubit operators and with the bosoniztion, the system can be studied using circuit quantum electrodynamics (C-QED) based quantum computation \cite {Adrian2024}.
    \item We induced representations that are stability subgroups generated by a single operator. So, they are the center (Casimir operators) of the algebra or the maximal torus of the algebra. The stability points are eigen vectors with the eigen value $1$. In essence, we have constructed IRRs with highest weight ($\lambda = 1$) with respect the Cartan subalgebras thar are generated by a single operator. This connection is significant as several Lie algebraic representation approaches are cast in terms of highest weight vectors while reducing the symmetries. This connection enables to go other way around by considering the corresponding groups and induce representations of larger symmetries.
    %\item The fermionic induced representation can be implemented using fermionic quantum computing approaches \cite {Ott202}.
    %\item Our rigorous treatment of induced representations in the context of SUSY for compact groups can be extended to locally compact groups such as the inhomogeneous Lorentz to fully incorporate relativistic effects.
    \item The q-deformed generalization of the above construction is rather straight forward and so we omitted the details.
\end{enumerate}

\section {Summary and conclusions}
We approached bosonization of SUSY using the tools of group representation theory. We illustrated the techniques of Mackey machinery for SUSY WZQM system by inducing representations from the fermionic to bosonic and the vice verza. To illustrate the power of bosonization, we constructed a tower of SUSY systems. We connected our induced representation to highest weight representation with $\lambda = 1$ and the Pauli subgroup with stability states serving as the maximal torus or the Cartan subalgebra. This connection opens up the way to lift the widely available representation results on reductive Lie algebraic methods on compact Lie groups to go in the opposite direction, that is constructing a larger symmetry from smaller ones. In the next step, we will fermionize an Adinkra and develop circuits for its implementation suitable for a fermionic quantum computer.
\section {Appendix}

{\bf Acknowledgment.} 
The research of S.J.G. is currently supported in part by the Clark Leadership Chair in Science endowment at the University of Maryland - College Park and the Army Research Office under Contract WNW911NF2520117.

\bibliographystyle{amsplain}

\end{document}